\documentclass[nofootinbib,prl,showkeys,showpacs]{revtex4-1}
\usepackage{graphicx}
\usepackage{natbib}
\usepackage{bm}
\usepackage{amssymb}
\usepackage{amsmath}
\usepackage{euscript}
\usepackage{mathrsfs}
\usepackage{esint}
\usepackage{slashed}

\begin{document}

\title{$1/f$ noise and quantum indeterminacy}

\author{Kirill A. Kazakov}

\affiliation{Moscow State University, 119991, Moscow, Russian Federation}
\begin{abstract}
An approach to the problem of $1/f$ voltage noise in conductors is developed based on an uncertainty relation for the Fourier-transformed signal. The quantum indeterminacy caused by non-commutativity of the observables at different times makes the voltage autocovariance ambiguous, but the power spectrum of fluctuations remains well-defined. It is shown that a lower bound on the power spectrum exists, which is related to the antisymmetric part of the voltage correlation function. Using the Schwinger-Keldysh method, this bound is calculated explicitly in the case of unpolarized charge carriers with a parabolic dispersion, and is found to have a $1/f$ low-frequency asymptotic. A comparison with the $1/f$-noise measurements in InGaAs quantum wells is made which shows that the observed noise levels are only a few times higher than the bound established.
\end{abstract}

\maketitle

As is well-known from the experiment, power spectra of the voltage fluctuations in all conducting media exhibit a universal low-frequency behavior: for sufficiently small frequencies $f,$ the power spectral density $S(f)\sim 1/f^{\gamma},$ where the frequency exponent $\gamma$ is around unity \cite{buckingham,bell1980,raychaudhuri}. These ubiquitous fluctuations are often called simply $1/f$ noise. It has been detected at frequencies as high as $10^6$Hz down to $10^{-6.3}$Hz, with no indication of a low-frequency spectrum flattening \cite{rollin1953,caloyannides}. There are various physical processes producing voltage noise: charge carrier trapping-detrapping, motion of dislocations, conductance fluctuations caused by the temperature fluctuations, {\it etc.} The so-called $1/f$-noise problem can be broadly formulated as a difficulty to relate the $1/f$-spectrum to any of these conventional noise sources. In fact, it is hard to indicate a physical process, say, in a crystal of pure copper, that would be characterized by frequencies much lower than one Hertz. What makes this problem a true mystery is that the observed absence of a low-frequency cutoff is in apparent conflict with finiteness of the voltage variance. In fact, a direct consequence of the familiar Wiener-Khinchin relation \cite{wiener,khinchin} is that this variance is equal to $2\int_{0}^{\infty} {\rm d}f S(f)=\infty.$ Two possible courses of action in this situation are found in the literature. It was suggested that the logarithmic divergence of the integral at $f=0$ is immaterial in view of the existence of a natural low-frequency cutoff, $f_0,$ -- the inverse of Universe lifetime, which bounds the total power to reasonably moderate values \cite{flinn1968}. However, not saying that this argument leads to an uncomfortable conclusion that the voltage across a small isolated conductor ought to behave differently in a static universe ($f_0=0$), it does not resolve the problem for $\gamma>1,$ because $\int {\rm d}f S(f) \sim 1/f^{\gamma-1}$ will be unacceptably large for many actual spectra continued down to $f_0.$ Since the Wiener-Khinchin theorem is valid only for processes characterized by stationary autocovariance, the other way to resolve the conflict is to allow for non-stationarity of the process generating $1/f$ noise. A number of mathematical models has been proposed in this direction which endow electric circuits with certain properties that allow them to act as filters with the response function being a fractional power of frequency, or alternatively, to behave as strange attractors (a survey of this approach can be found in Ref.~\cite{bell1980}, a more recent development, in Refs.~\cite{leibovich2015,dechant2015}). A major issue with these models when applied to the voltage noise is that materials used in microelectronics and related areas do not exhibit desired properties. The purpose of this Letter is to point at an essentially different possibility to resolve the problem, naturally suggested by the quantum theory. As is well known, this theory puts certain restrictions on pairwise measurability of physical quantities, which are expressed by the uncertainty relations. It will be shown that this quantum indeterminacy obstructs applicability of the Wiener-Khinchin theorem, and puts a nontrivial lower bound on the power spectrum. This bound will be explicitly calculated and demonstrated to have the $1/f$ low-frequency asymptotic.

Consider a sample with a constant electric current through it supplied by two leads attached to the sample. For simplicity, the sample material will be assumed macroscopically homogeneous, and so will be the electric field, $\bm{E},$ established inside the sample. Let the voltage across the sample be measured by means of two voltage probes which may or may not coincide with the current leads. Also for simplicity, the probes will be considered pointlike, $\bm{x}_1,\bm{x}_2$ denoting their position. A voltage $U(t,\bm{x}_1,\bm{x}_2)$ measured between the probes at time $t$ is the sum of a constant bias $U_0(\bm{x}_1,\bm{x}_2)$ and a fluctuation, or noise, $\Delta U(t,\bm{x}_1,\bm{x}_2),$
\begin{eqnarray}\label{voltagedecomp}
U(t,\bm{x}_1,\bm{x}_2) = U_0(\bm{x}_1,\bm{x}_2) + \Delta U(t,\bm{x}_1,\bm{x}_2).
\end{eqnarray}
Define a Fourier transform of $\Delta U(t,\bm{x}_1,\bm{x}_2)$ measured during time $t_m$ (omitting, for brevity, the parameters $\bm{x}_1,\bm{x}_2$)
\begin{eqnarray}\label{fourier}
\Delta U_s(\omega) = \int_{0}^{t_m}{\rm d}t\, \Delta U(t)\sin(\omega t),
\quad \Delta U_c(\omega) = \int_{0}^{t_m}{\rm d}t\, \Delta U(t)\cos(\omega t).
\end{eqnarray} After this measurement has been repeated many times, the main quantity of interest -- the noise power spectral density -- is obtained as
\begin{eqnarray}\label{power}
S(f) = \lim\limits_{t_m\to \infty}\frac{1}{t_m}\left\langle \left(\Delta U_s(\omega)\right)^2 + \left(\Delta U_c(\omega)\right)^2\right\rangle, \quad \omega = 2\pi f,
\end{eqnarray} where the angular brackets denote averaging over the set of signal samplings, and $t_m\to\infty$ designates the condition that the measurement time be much greater than the given frequency inverse. In the context of the Wiener-Khinchin theorem, $\Delta U(t)$ is a random variable which is classical in that it takes on a real value at any given $t$ (a voltmeter reading). The theorem states that if the autocovariance $\langle \Delta U(t)\Delta U(t+\tau)\rangle \equiv C(\tau)$ is a function of the lag $\tau$ only, then
\begin{eqnarray}\label{wiener}
C(\tau) = \int_{-\infty}^{+\infty}{\rm d}fS(f){\rm e}^{{\rm i}2\pi f\tau}\,.
\end{eqnarray}
In the case of a quantum underlying process, in order to evaluate the function $C(\tau),$ one needs to construct a quantum-mechanical operator that would represent the product $\Delta U(t)\Delta U(t+\tau).$ From the quantum theory standpoint, the voltage $U(t)$ is an observable to which there corresponds a Hermitian (Heisenberg) operator $\widehat{U}(t).$ For a given bias $U_0(\bm{x},\bm{x}')\equiv U_0$, this defines another observable -- the voltage fluctuation -- by virtue of Eq.~(\ref{voltagedecomp}),
\begin{eqnarray}\label{voltagedecomp1}
\widehat{\Delta U}(t) = \widehat{U}(t) - U_0.
\end{eqnarray}
But one cannot construct an observable for the product $\Delta U(t)\Delta U(t+\tau)$ by multiplying $\widehat{\Delta U}(t)$ and $\widehat{\Delta U}(t+\tau),$ because these operators do not commute with each other, so that their multiplication is ambiguous. As is well known, the physical meaning of this ambiguity is that $\Delta U$ cannot have definite values at two instants $t,t+\tau$ in any state with a nonzero expectation value of the commutator $[\widehat{\Delta U}(t),\widehat{\Delta U}(t+\tau)].$ Put somewhat differently, it is impossible to obtain definite values of $\Delta U$ at two different instants without changing the system state: even if the state vector of the system ``conducting sample plus electromagnetic field'' was changed negligibly during the first measurement with a definite outcome, it will necessarily undergo a finite change during a later measurement that results in a definite $\Delta U.$ It follows that each sampling of $\Delta U$ during the time $t_m$ is accompanied by a continuous alteration of the system state in a way consistent with the uncertainty principle: a higher measurement precision entails larger system state variations, hence larger voltage fluctuations between successive measurements. Therefore, the voltage autocovariance will depend on the way the system state varies during the measurement, in particular, it will be a function of both arguments $t$ and $(t+\tau).$ An alternative we thus face is that {\it either, for a fixed state, the voltage fluctuation is not defined for all times in principle, or $\left\langle\widehat{\Delta U}(t)\widehat{\Delta U}(t+\tau)\right\rangle$ is $t$-dependent.} But since the Wiener-Khinchin theorem assumes both that the process $\Delta U(t),$ though random, is defined for all $t$ and is stationary, it does not apply in either case.

It is important, on the other hand, that the quantum-mechanical description of the power spectrum is free of this problem. To see this, we first note that $\widehat{\Delta U}(t)$ uniquely defines two other Hermitian operators $\widehat{\Delta U_s}(\omega),$ $\widehat{\Delta U_c}(\omega)$ according to Eq.~(\ref{fourier})
\begin{eqnarray}\label{fourier1}
\widehat{\Delta U_s}(\omega) = \int_{0}^{t_m}{\rm d}t\, \widehat{\Delta U}(t)\sin(\omega t),
\quad \widehat{\Delta U_c}(\omega) = \int_{0}^{t_m}{\rm d}t\, \widehat{\Delta U}(t)\cos(\omega t),
\end{eqnarray} and that the squares of these are also uniquely defined despite the non-commutativity of $\widehat{\Delta U}(t)$ at different $t$s. Specifically, they are expressed via the symmetrized product of $\widehat{\Delta U}(t)$
\begin{eqnarray}
\left(\widehat{\Delta U_s}(\omega)\right)^2 &=& \iint_{0}^{t_m}{\rm d}t{\rm d}t'\, \widehat{\Delta U}(t) \widehat{\Delta U}(t')\sin(\omega t)\sin(\omega t') \nonumber\\ &=& \iint_{0}^{t_m}{\rm d}t{\rm d}t'\, \frac{1}{2}\left(\widehat{\Delta U}(t)\widehat{\Delta U}(t')+\widehat{\Delta U}(t')\widehat{\Delta U}(t)\right)\sin(\omega t)\sin(\omega t'),
\end{eqnarray} and similarly for $\left(\widehat{\Delta U_c}(\omega)\right)^2.$ And second, the definition of $S(f)$ according to Eq.~(\ref{power})  exactly corresponds to the usual quantum-mechanical formula for calculating averages,
\begin{eqnarray}\label{powerq1}
S(f) = \lim\limits_{t_m\to \infty}\frac{1}{t_m}\left\{\left\langle  \left(\widehat{\Delta U_s}(\omega)\right)^2\right\rangle + \left\langle\left(\widehat{\Delta U_c}(\omega)\right)^2\right\rangle\right\},
\end{eqnarray} where $\left\langle \hat{f} \right\rangle = {\rm tr}\left( \hat{\rho}\hat{f}\right),$ $\hat{\rho}$ being the system density matrix. In fact, it is the definition of an observable $\hat{f}$ that ${\rm tr}\left( \hat{\rho}\hat{f}\right)$ equals the mean of a set of $f$-values obtained in a series of measurements in the given state $\hat{\rho}$. Here, the non-commutativity of $\widehat{\Delta U}(t)$ at different $t$s shows itself as a non-commutativity of the two operators $\left(\widehat{\Delta U_s}(\omega)\right)^2$ and $\left(\widehat{\Delta U_c}(\omega)\right)^2$, which, however, causes presently no ambiguity, because the average of the sum is equal to the sum of averages. Thus, we arrive at the conclusion that in contrast to the autocovariance, the power spectrum of voltage fluctuations can be self-consistently computed using the formula (\ref{powerq1}).

Now, the question is if this quantum indeterminacy puts any nontrivial bound on the level of voltage fluctuations. Such bound is naturally expected to be independent of the measurement process specifics. Therefore, we exclude the measuring device from consideration by assuming that $\hat{\rho}$ describes a fixed state of the system ``conducting sample plus electromagnetic field.'' An uncertainty relation for the two observables $\Delta U_s(\omega), \Delta U_c(\omega)$ then reads
\begin{eqnarray}\label{indeterm}
\left\langle \left(\widehat{\Delta U_s}(\omega)\right)^2\right\rangle \left\langle \left(\widehat{\Delta U_c}(\omega)\right)^2\right\rangle \geqslant \frac{1}{4}\left| \left\langle \left[\widehat{\Delta U_s}(\omega),\widehat{\Delta U_c}(\omega)\right]\right\rangle\right|^2.
\end{eqnarray} It readily follows from Eqs.~(\ref{powerq1}), (\ref{indeterm}) that the minimum of $S(f)$ is
\begin{eqnarray}\label{powerf}
S_F(f)&=&\lim\limits_{t_m\to \infty}\frac{1}{t_m}\left| \left\langle \left[\widehat{\Delta U_s}(\omega),\widehat{\Delta U_c}(\omega)\right]\right\rangle\right|.
\end{eqnarray} This quantity represents what can be called a {\it fundamental noise}. An important observation to be made is that the expectation value of the commutator on the right hand side of Eq.~(\ref{powerf}) is an {\it odd} function of frequency, for $\widehat{\Delta U_s}(\omega)$ is odd, whereas $\widehat{\Delta U_c}(\omega)$ is even in $\omega$. More specifically, using Eq.~(\ref{fourier1}) and interchanging the integration variables gives
\begin{eqnarray}\label{commutator}
\left\langle\left[\widehat{\Delta U_s}(\omega),\widehat{\Delta U_c}(\omega)\right]\right\rangle &=& \iint_{0}^{t_m}{\rm d}t{\rm d}t'\, \left\langle[\widehat{\Delta U}(t),\widehat{\Delta U}(t')]\right\rangle\sin(\omega t)\cos(\omega t') \nonumber\\ &=& \iint_{0}^{t_m}{\rm d}t{\rm d}t'\, \left\langle\widehat{\Delta U}(t)\widehat{\Delta U}(t')\right\rangle\sin(\omega (t - t')).
\end{eqnarray} It follows that $S_F(f)$ is determined by the Fourier transform of the contribution to $\left\langle\widehat{\Delta U}(t)\widehat{\Delta U}(t')\right\rangle\equiv S(t-t')$ which is antisymmetric under the interchange $t\leftrightarrow t'.$ Thus, when computing this expectation value, one can safely use the Fourier  techniques to extract its low-frequency asymptotic despite anticipated zero-frequency singularity, for unlike an even-frequency function $1/|\omega|,$ Fourier transform of $1/\omega$ does exist. On the other hand, $S(f)$ is determined by the symmetric part of $S(t-t'),$
$$\left\langle  \left(\widehat{\Delta U_s}(\omega)\right)^2\right\rangle + \left\langle\left(\widehat{\Delta U_c}(\omega)\right)^2\right\rangle = \iint_{0}^{t_m}{\rm d}t{\rm d}t'\, \left\langle\widehat{\Delta U}(t)\widehat{\Delta U}(t')\right\rangle\cos(\omega (t - t')),$$
and this part lacks Fourier decomposition when $S(f)\sim 1/|\omega|.$ We see that while $S_F(f)$ is derived from a function admitting Fourier decomposition, Fourier transform of $S_F(f)$ itself may not exist. In other words, the right hand side of Eq.~(\ref{wiener}) does not have to be well-defined, and as we have seen, neither does its left hand side.

Regarding the relationship between the power spectrum and autocovariance, one must also note the following. Consider the quantity
$$\Sigma(f) = \lim\limits_{t_m\to \infty}\frac{1}{t_m}\iint_{0}^{t_m}{\rm d}t{\rm d}t'\, S(t-t') {\rm e}^{{\rm i}\omega (t - t')}.$$ On applying Euler's formula to the exponent in the integrand, the contribution of cosine to $\Sigma(f)$ is just $S(f),$ whereas the modulus of the other contribution is $S_F(f).$ Using integration by parts, this can be reduced to
\begin{eqnarray}\label{sigma}
\Sigma(f) = \lim\limits_{t_m\to \infty}\left\{\int_{-t_m}^{t_m}{\rm d}\tau S(\tau) {\rm e}^{{\rm i}\omega\tau} - \frac{1}{t_m}\int_{-t_m}^{t_m}{\rm d}\tau |\tau|S(\tau) {\rm e}^{{\rm i}\omega\tau}\right\}.
\end{eqnarray} If one assumes that $S(\tau)$ sufficiently rapidly decreases as $|\tau|\to \infty,$ the second term yields zero in the limit $t_m\to\infty,$ and Eq.~(\ref{sigma}) takes the same form as the conventional Wiener-Khinchin relation for processes with a well-defined autocovariance. But the $1/f$-noise problem is just the case where this assumption does not hold. The $1/|\omega|$ asymptotic of the power spectrum suggests that at large $|\tau|$'s, $S(\tau)$ contains a term $\sim \ln|\tau|.$ It is not difficult to check that
\begin{eqnarray}\label{integrals}
\int_{-t_m}^{t_m}{\rm d}\tau \ln|\tau| {\rm e}^{{\rm i}\omega\tau} &=& -\frac{\pi}{|\omega|} - 2\ln t_m\frac{\sin(\omega t_m)}{\omega}  + O(1/t_m), \nonumber \\
\frac{1}{t_m}\int_{-t_m}^{t_m}{\rm d}\tau |\tau|\ln|\tau| {\rm e}^{{\rm i}\omega\tau} &=& - 2\ln t_m\frac{\sin(\omega t_m)}{\omega}  + O(1/t_m).\end{eqnarray} It is seen that despite these expressions diverge in the limit $t_m\to\infty,$ their difference converges to $-\pi/|\omega|.$ It follows that the right hand side of Eq.~(\ref{sigma}) is well-defined for a function $S(\tau) = \ln[a + (\tau/\tau_0)^2]$ with constant $a,\tau_0>0,$ and that $\Sigma(f) \to -1/|f|$ for $|f|\ll 1/\tau_0.$ This simple example well illustrates the above statement that the power spectrum may exist even for processes for which $S(t-t')$ does not admit Fourier decomposition, and that the presence of a $1/f$-term in its low-frequency asymptotic does not contradict finiteness of the voltage variance. In fact, in the present example, $S(0) = \ln a$ can be any real number despite the ``total power'' $\int_{-\infty}^{\infty}{\rm d}f\Sigma(f)$ is infinite.

Next, the low-frequency asymptotic of $S_F(f)$ will be calculated explicitly. The calculation will be simplified as much as possible: we assume that there is a single species of free-like charge carriers with charge $e$ and effective mass $m$; they are taken non-relativistic unpolarized fermions, and all spin indices are accordingly suppressed; all finite-temperature effects (due to the photon heat bath, in particular) are neglected; the contributions independent of the charge-carrier velocity are only taken into account. The expectation value $\left\langle\widehat{\Delta U}(t)\widehat{\Delta U}(t')\right\rangle$ will be computed using the Schwinger-Keldysh technique \cite{schwinger,keldysh} according to which it
can be written as
\begin{eqnarray}\label{expectation2}
\left\langle\widehat{\Delta U}(t)\widehat{\Delta U}(t')\right\rangle = {\rm tr}\left(\hat{\rho}\,\EuScript{T}_C\widehat{\Delta u}^{(2)}(t)\widehat{\Delta u}^{(1)}(t')\exp\left\{-{\rm i}\int_{C}{\rm d}t\,\hat{w}(t)\right\}\right),
\end{eqnarray}
where the lower-case letters denote operators in the interaction picture; the so-called Schwinger-Keldysh time contour $C$ runs from $t=-\infty$ to $t=+\infty$, and then back to $t=-\infty,$ with all time instants on the forward branch designated with a superscript $(1)$ and treated as being in the past with respect to any time instant on the backward branch designated with a superscript $(2).$ $\EuScript{T}_C$ denotes operator ordering along this contour. The Hamiltonian of the charge carrier--electromagnetic field interaction reads
\begin{eqnarray}\label{hamiltonian}
\hat{w}(t) = \int {\rm d}^3\bm{x}\left[e\hat{a}_0(x) + \frac{e^2}{2mc^2}\hat{\bm{a}}^2(x) \right]\hat{\phi}^{\dagger}(x) \hat{\phi}(x), \quad x = (ct,\bm{x}),
\end{eqnarray} where $\hat{\phi}$ is the fermionic field, and $(\hat{a}_0,\hat{\bm{a}})$ is the electromagnetic 4-potential. The electromagnetic field propagator will be used in the Feynman gauge, according to which the direct Coulomb interaction of charge carriers was excluded from the interaction Hamiltonian \cite{weinberg,landsman}.
Since the propagator is diagonal in this gauge, and the charge carriers are non-relativistic, one can write $\widehat{\Delta u}(t) = \hat{a}_{0}(t,\bm{x}_1)-\hat{a}_{0}(t,\bm{x}_2) - U_0,$ and replace $\hat{\bm{a}}$ in Eq.~(\ref{hamiltonian}) by the classical field
$${\bm a}(x) = {\rm i}c\bm{E}\frac{{\rm e}^{{\rm i\lambda t}} - 1}{\lambda}\,, \quad \lambda \to 0.$$ This form is chosen in order to obtain Fourier decomposition for the antisymmetric part of $S(t-t').$   Switching for a moment to the units $\hbar=c=1,$ the lowest-order nontrivial contribution to the right hand side of Eq.~(\ref{expectation2}) reads
\begin{eqnarray}\label{lowest}
&&S(\tau) = \frac{{\rm i}(4\pi e^2)^2\bm{E}^2}{m}\frac{\partial^2}{\partial\lambda\partial\lambda'}\int\frac{{\rm d}^4 k}{(2\pi)^4}\frac{{\rm d}^3 \bm{q}}{(2\pi)^3}\frac{{\rm d}^3 \bm{q}'}{(2\pi)^3}\varrho(\bm{q},\bm{q}')\left[{\rm e}^{{\rm i}\bm{k}\cdot(\bm{x}_1 - \bm{x}_2)} - 1\right] {\rm e}^{{\rm i}(\bm{q}-\bm{q}')\cdot\bm{x}_1 }\nonumber\\&&\times\left\{{\rm e}^{-{\rm i}k^0\tau}G^{(11)}(q-q'+k+k')D^{(11)}(q+k+k')D^{(11)}(q+k)G^{(12)}(k)
\right.\nonumber\\ &&\left.\left. - {\rm e}^{{\rm i}k^0\tau} G^{(22)}(q-q'+k+k')D^{(22)}(q+k+k')D^{(22)}(q+k)G^{(21)}(k)\right\}\right|_{\lambda=\lambda'=0} + (\bm{x}_1 \leftrightarrow\bm{x}_2), \nonumber
\end{eqnarray} where $k' = (\lambda+\lambda',\bm{0}),$ $G$ and $D$ are the momentum-space Schwinger-Keldysh propagators of the electromagnetic and charge carrier fields, respectively,
\begin{eqnarray}\label{photonprop}
G^{(11)}(k) &=& \frac{{\rm i}}{k^2 + {\rm i}0} = [G^{(22)}(k)]^*, \quad G^{(12)}(k) = 2\pi\theta(-k^0)\delta(k^2), \quad G^{(21)}(k) = 2\pi \theta(k^0)\delta(k^2),\nonumber \\
D^{(11)}(q) &=& \frac{{\rm i}}{q^0 - \varepsilon_{\bm{q}} + {\rm i}0} = [D^{(22)}(q)]^*, \quad k=(k^0,\bm{k}), \quad q=(\varepsilon_{\bm{q}},\bm{q}),
\end{eqnarray}  and the step function $\theta(k^0)=0$ for $k^0 \leqslant 0$, $\theta(k^0)=0$ for $k^0>0$; since the photon heat bath effects are neglected, $G$'s are purely vacuum; accordingly, the system density matrix $\hat{\rho}$ is reduced to that of the charge carriers, $\varrho(\bm{q},\bm{q}')$ denoting its momentum-space representation. The leading low-frequency term comes from $\lambda$-differentiation of the charge-carrier propagator $D^{(11)}(q+k+k')$, because it approaches its pole as $k^0\to 0.$ Since $|\bm{k}| = |k^0|,$ one has $\varepsilon_{\bm{q+k}} = \varepsilon_{\bm{q}} + |k^0|O(|\bm{q}|/m),$ and therefore, in the non-relativistic limit, $D^{(11)}(q+k+k') = {\rm i}/(q^0 + k^0 + \lambda+\lambda' - \varepsilon_{\bm{q+k}})\approx {\rm i}/(k^0+\lambda+\lambda').$ Expanding also ${\rm e}^{{\rm i}\bm{k}\cdot(\bm{x}_1 - \bm{x}_2)}$ to the second order, and performing integration over $\bm{k}$ gives
\begin{eqnarray}\label{lowest}
&&S(\tau)= \frac{16e^4\bm{E}^2(\bm{x}_1 - \bm{x}_2)^2}{3m}\int_{0}^{\infty}{\rm d}k^0\frac{{\rm e}^{{\rm i}k^0\tau}}{k^0}\int\frac{{\rm d}^3 \bm{q}}{(2\pi)^3}\frac{{\rm d}^3 \bm{q}'}{(2\pi)^3}\varrho(\bm{q},\bm{q}')\frac{{\rm e}^{{\rm i}(\bm{q}-\bm{q}')\cdot\bm{x}_1 } + {\rm e}^{{\rm i}(\bm{q}-\bm{q}')\cdot\bm{x}_2 }}{(\bm{q} - \bm{q}')^2}\,.
\end{eqnarray} As anticipated, we have a convergent frequency integral for the antisymmetric part of $S(t-t')$, namely, $\int_{0}^{\infty}{\rm d}k^0\sin(k^0\tau)/k^0 = \pi \tau/(2|\tau|).$ The obtained expression for the antisymmetric part is to be substituted into the right hand side of Eq.~(\ref{sigma}). Using the formulas
\begin{eqnarray}
\int_{-t_m}^{t_m}{\rm d}\tau \frac{\tau}{|\tau|} {\rm e}^{{\rm i}\omega\tau} &=& \frac{2{\rm i}}{\omega}[1- \cos(\omega t_m)], \quad
\frac{1}{t_m}\int_{-t_m}^{t_m}{\rm d}\tau \tau{\rm e}^{{\rm i}\omega\tau} = - \frac{2{\rm i}}{\omega}\cos(\omega t_m) + O(1/t_m), \nonumber
\end{eqnarray} we see that, just like in the above example with $\ln|\tau|$ in the symmetric part of $S(\tau),$ neither of the two integrals in Eq.~(\ref{sigma}) converge, but their difference has a well-defined limit as $t_m\to \infty.$

In practice, the voltage probes are usually aligned parallel to $\bm{E},$ in which case $\bm{E}^2(\bm{x}_1 - \bm{x}_2)^2 = U^2_0.$ The remaining momentum integrals are conveniently evaluated by expressing $\varrho(\bm{q},\bm{q}')$ via the mixed position-momentum distribution function $f(\bm{r},\bm{Q}),$
\begin{eqnarray}\label{mixed}
\varrho\left(\bm{Q}-\frac{\bm{p}}{2},\bm{Q}+\frac{\bm{p}}{2}\right) = \int_{\Omega} {\rm d}^3\bm{r}{\rm e}^{{\rm i}\bm{p}\cdot\bm{r}}f(\bm{r},\bm{Q}),
\end{eqnarray} $\Omega$ denoting the sample volume. The integral over $\bm{p}=\bm{q}'-\bm{q}$ in Eq.~(\ref{lowest}) is then a Fourier decomposition of the Coulomb potential. Probability distribution for the particle position in the sample can be obtained by integrating
$f(\bm{r},\bm{Q})$ over all momenta $\bm{Q} = (\bm{q}+\bm{q}')/2.$ In view of the assumed macroscopic sample homogeneity, $\int{\rm d}^3 \bm{Q}f(\bm{r},\bm{Q})$ is position-independent within the sample (vanishing outside of it). Thus, restoring ordinary units, the power spectrum takes the form
\begin{eqnarray}\label{powerfinal}
S_F(f) = \frac{\varkappa U^2_0}{|f|}\,, \quad \varkappa \equiv \frac{2e^4 g}{\pi m\hbar c^3}\,,
\end{eqnarray} where $g$ is a geometrical factor $$g = \frac{1}{3\Omega}\int\limits_{\Omega}{\rm d}^3\bm{r}\left(\frac{1}{|\bm{r}-\bm{x}_1|} + \frac{1}{|\bm{r}-\bm{x}_2|}\right).$$

Regarding this formula, it should be noted that while the picture of free-like charge carriers could be used to find the lowest-order term of $S_F$ with respect to $\bm{E},$ the higher-order contributions cannot be calculated in the same way, because of the particle collisions. The easiest way to see this is to consider the effect of collisions as an $O(e)$ modification of the external field $\bm{E}$ that acts on the given charge carrier. The fields describing particle interactions are much stronger than $\bm{E},$ rapidly varying in space and time, but the mean value of the net field vanishes. Therefore, it is the field variance that affects $S_F\sim\bm{E}^2,$ so that account of the particle collisions would add terms $O(e^6)$ to the right hand side of Eq.~(\ref{powerfinal}). On the other hand, since $S_F$ must be even in $\bm{E},$ the next term of its expansion in powers of the external electric field is $\sim(\bm{E}^2)^2,$ hence it is also $O(e^6),$ and so keeping this term in the present picture would not be legitimate.

To give an idea of how far the observed noise levels are from the minimum value (\ref{powerfinal}), take for example Ref.~\cite{chenaud2016} reporting noise measurements in InGaAs quantum wells of significantly different sizes. The charge carriers in this case are electrons ($n$) and (light) holes ($p$); their effective masses depend on the sample thickness, composition {\it etc.}, but for a rough estimate it will suffice to take $m_n = 0.06m_0$, $m_p = 0.09m_0,$ respectively, where $m_0$ is the free electron mass. The sample dimensions and experimental results taken from figures~4--7 of \cite{chenaud2016} are compared in Table~\ref{table} with the values computed according to Eq.~(\ref{powerfinal}).
\begin{table}
\begin{tabular}{ccccc||c||c}
\hline\hline
{\rm sample}
  & \hspace{0,1cm} $w,$ $\mu$m \hspace{0,1cm}
  & \hspace{0,1cm} $l,$ $\mu$m \hspace{0,1cm}
  & \hspace{0,1cm} $a,$ nm \hspace{0,1cm}
  & $g,$ cm$^{-1}$
  & \hspace{0,5cm} $\varkappa_{\rm th}$ \hspace{0,5cm}
  &  \hspace{0,5cm} $\varkappa_{\rm exp}$  \hspace{0,5cm} \\
\hline
V1& 1 & 2.2 & 10 & 9630 & $3.5\cdot 10^{-10}$ & $1.75\cdot 10^{-9}$\\
V1.5& 1.5 & 3.3 & 10 & 6420 & $2.3\cdot 10^{-10}$ & $4.5\cdot 10^{-10}$\\
V2& 2 & 4 & 10 & 5140 & $1.9\cdot 10^{-10}$ & $3.1\cdot 10^{-10}$\\
V5& 5 & 20 & 20 & 1260 & $4.6\cdot 10^{-11}$ & $2.9\cdot 10^{-9}$ \\
V80& 80 & 310 & 20 &  80  & $1.9\cdot 10^{-12}$ & $4.1\cdot 10^{-12}$\\
\hline\hline
\end{tabular}
\caption{Comparison of the measured ($\varkappa_{\rm exp}$) in Ref.~\cite{chenaud2016} and calculated ($\varkappa_{\rm th}$) according to Eq.~(\ref{powerfinal}) noise magnitude in various InGaAs samples. Also given are the sample width ($w$), length ($l$) and thickness ($a$).} \label{table}
\end{table}
It is seen that with one exception, the measured noise levels are only a few times higher than the lower bound set by the quantum indeterminacy. The exceptionally large value of $\varkappa_{\rm exp}$ detected in sample V5 is actually related to deviation of the frequency exponent from unity, which is noticeable in this case. Discussion of this issue as well as account of the heat-bath contributions and further details of the calculations performed will be given elsewhere.

\pagebreak

\end{document}